\documentclass[a4paper,10pt]{article}
\usepackage{caption}
\usepackage{jinstpub} 
\usepackage{lineno}
\usepackage{graphicx}
\usepackage{siunitx}
\usepackage{orcidlink}
\usepackage[]{changes}

\definechangesauthor[name={Simone}, color=red]{A}

\newif\ifdraft
\drafttrue

\title{
Machine-Learning–Enhanced Event Selection for BEGe Detectors in the VIP Experiment
}

\author{S. Manti$^{1,*,\dagger}$\orcidlink{0000-0003-3770-0863}}
\author{J.  Yip$^{2,\dagger}$\orcidlink{0009-0001-6947-4330}}
\author{M. Bazzi$^{1}$\orcidlink{0000-0002-1699-7138}}
\author{N. Bortolotti$^{1,3}$\orcidlink{0000-0002-9344-3657}}
\author{M. Bragadireanu$^{4}$\orcidlink{0009-0001-5217-6003}}
\author{I. Carnevali$^{1}$}
\author{A. Clozza$^{1}$\orcidlink{0000-0003-2133-1725}}
\author{L. De Paolis$^{1}$\orcidlink{0000-0002-4203-9902}}
\author{R. Del Grande$^{5,1}$\orcidlink{0000-0002-7599-2716}}
\author{C. Guaraldo$^{1,\ddagger}$\orcidlink{0000-0002-8923-3438}}
\author{M. Iliescu$^{1}$\orcidlink{0009-0003-3859-5679}}
\author{M. Laubenstein$^{6}$\orcidlink{0000-0001-5390-4343}}
\author{J. Marton$^{7,8}$\orcidlink{0009-0003-1912-285X}}
\author{F. Nola$^{1,9,10}$\orcidlink{0000-0001-9249-9547}}
\author{K. Pischicchia$^{3,1}$\orcidlink{0000-0001-6879-452X}}
\author{A. Porcelli$^{11,12}$\orcidlink{0000-0002-3220-6295}}
\author{A. Scordo$^{1}$\orcidlink{0000-0002-7703-7050}}
\author{F. Sgaramella$^{1}$\orcidlink{0000-0002-0011-8864}}
\author{D. Sirghi$^{3,1,4}$\orcidlink{0009-0002-7486-025X}}
\author{F. Sirghi$^{1,4}$\orcidlink{0000-0002-6143-3200}}
\author{J. Zmeskal$^{7,\ddagger}$\orcidlink{0000-0003-0815-0639}}
\author{C. Curceanu$^{1,4}$\orcidlink{0000-0002-1990-0127}}

\affiliation{$^{1}$ Laboratori Nazionali di Frascati INFN, Frascati, Italy}
\affiliation{$^{2}$ University of California, Berkeley, CA 94720, USA}
\affiliation{$^{3}$ Centro Ricerche Enrico Fermi, Museo Storico della Fisica e Centro Studi e Ricerche "Enrico Fermi", Roma, Italy}
\affiliation{$^{4}$ IFIN-HH, Institutul National pentru Fizica si Inginerie Nucleara Horia Hulubei, 30 Reactorului, 077125, Magurele, Romania}
\affiliation{$^{5}$ Faculty of Nuclear Sciences and Physical Engineering, Czech Technical University in Prague, Břehovà 7, 115 19, Prague, Czech Republic}
\affiliation{$^{6}$ Laboratori Nazionali del Gran Sasso, Istituto Nazionale di Fisica Nucleare, Via G. Acitelli 22, 67100, Assergi, Italy}
\affiliation{$^{7}$ Stefan Meyer Institute for Subatomic Physics, Vienna, Austria}
\affiliation{$^{8}$ Atominstitut, Technische Universität Wien, Stadionallee 2, 1020 Vienna, Austria}
\affiliation{$^{9}$ Istituto Nazionale di Fisica Nucleare, Complesso Universitario di Monte S. Angelo, Via Cintia - I-80126 Napoli, Italy}
\affiliation{$^{10}$ Dipartimento di Matematica e Fisica, Università degli Studi della Campania "Luigi Vanvitelli", viale Abramo Lincoln 5 - I-81100 Caserta, Italy}
\affiliation{$^{11}$ Centro de Investigacíon, Tecnología, Educacíon y Vinculacíon Astronómica, Universidad de Antofagasta, Avenida Angamos 601, 1240000, Antofagasta, Chile}
\affiliation{$^{12}$ Faculty of Physics, Astronomy, and Applied Computer Science, Jagiellonian University, Kraków, Poland}

\affiliation{$^*$ Corresponding Authors}
\affiliation{$^\dagger$ These authors contributed equally}
\affiliation{$^\ddagger$ Deceased}

\emailAdd{simone.manti@lnf.infn.it}
\emailAdd{jason17@berkeley.edu}

\abstract{
The VIP collaboration operates a Broad-Energy Germanium (BEGe) detector at the Gran Sasso National Laboratory to measure radiation in the few-keV to 100 keV range, aiming to search for spontaneous collapse-induced radiation and atomic transitions violating the Pauli Exclusion Principle. Here, we present a machine-learning–based upgrade for the BEGe detector of an event-selection strategy aimed at improving the efficiency in detecting low-energy events down to 10 keV. The method employs a denoising autoencoder to suppress electronic and microphonic noises and reconstruct pulse shapes, followed by a convolutional neural network that classifies waveforms as normal single-site or event with anomalies. The workflow was validated on a dataset comprising more than 20,000 waveforms recorded in 2021. The classifier achieves a receiver operating characteristic (ROC) curve with an area under the curve (AUC) of 0.99 and an accuracy of 95\%. Applying this procedure lowers the minimum detectable energy of the final spectrum to approximately 10 keV. It also yields a measurable enhancement in spectral quality, including an improvement of about 14\% in the signal-to-background ratio and improvement of the energy resolution for the characteristic Pb and Bi gamma lines. These developments enhance the sensitivity of the BEGe detector to rare low-energy signals and provide a scalable framework for future precision tests of quantum foundations in low-background environments.
}


\arxivnumber{1234.56789} 

\begin{document}
\maketitle
\flushbottom

\section{Introduction}
Experimental tests of quantum mechanics foundations require extreme sensitivity, as they target phenomena that, if observed, would occur at extremely small rates \cite{porcelliStrongestConstraintParastatistical2025}. Such rare-event searches provide a unique window into possible deviations from the standard quantum framework, complementing more conventional precision tests and opening opportunities to probe new physics at the interface between microscopic and macroscopic scales \cite{bassiModelsWavefunctionCollapse2013}.
An example of such tests is the search for violations of the Pauli Exclusion Principle \cite{pauliConnectionSpinStatistics1940}. PEP is one of the most fundamental principles of quantum theory, which underlies the structure of atoms, the stability of matter \cite{dysonStabilityMatter1967}, and even astrophysical objects like neutron stars \cite{glendenningSpecialGeneralRelativity2010}. Its validity is derived from the Spin–Statistics Theorem \cite{streaterPCTSpinStatistics2000}.
This theorem is intimately connected to locality and Lorentz covariance \cite{guidoAlgebraicSpinStatistics1995}; therefore, any violation of PEP would have profound implications, pointing to physics beyond the established local quantum field theoretical framework \cite{greenbergParticlesSmallViolations1991,greenbergTheoriesViolationStatistics2000,balachandranSpinStatisticsGroenewold2006}. From an experimental point of view, such violations could manifest as anomalous atomic transitions, into already double-occupied states, shifted from the standard spectral lines \cite{rambergExperimentalLimitSmall1990}. Establishing increasingly stringent limits on such effects is therefore a central goal of dedicated rare-event searches.
Another direction involves models of spontaneous wave-function collapse, which resolve the measurement problem by introducing additional physical mechanisms that suppress quantum superpositions at macroscopic scales. Prominent examples include the Continuous Spontaneous Localization (CSL) \cite{ghirardiUnifiedDynamicsMicroscopic1986,ghirardiContinuousspontaneousreductionModelInvolving1990} and the Diósi–Penrose (DP) \cite{diosiModelsUniversalReduction1989,penroseGravitysRoleQuantum1996}. Both predict the spontaneous emission of radiation, in particular X-rays, which can serve as experimental signatures of the individual collapse model, especially in the low-energy regime (10-100 keV) \cite{piscicchiaXRayEmissionAtomic2024b}.\newline
Detecting such rare and subtle signatures requires not only detectors with excellent energy resolution but also an extremely low-background environment. Underground laboratories play a crucial role in this context. The Gran Sasso National Laboratory (LNGS) of INFN provides a uniquely low-background environment for the study of rare events. Located beneath approximately \SI{1400}{m} of rock, a water equivalent (w.e.) to about \SI{3800}{\text{m.w.e.}}, the Gran Sasso overburden suppresses the cosmic muons by a factor of $10^6$, i.e. a flux of \SI{3.41e-4}{m^{-2}s^{-1}} \cite{belliniCosmicmuonFluxAnnual2012}. This substantial attenuation of the muon flux enables experiments at LNGS to achieve significantly reduced background levels. This makes LNGS one of the most favorable environments for rare-event searches, hosting experiments ranging from neutrino physics to dark matter and double-beta decay \cite{xenoncollaborationFirstDarkMatter2023,majoranacollaborationSearchNeutrinolessDouble$ensuremathbeta$2018,gerdacollaborationFinalResultsGERDA2020,ackermannGerdaExperimentSearch2013}.
Taking advantage of this low-background environment, the VIP (Violation of the Pauli Exclusion Principle) collaboration has progressively tightened the upper limits on PEP-violating searches. The original VIP experiment set an upper limit on the PEP violating probability for electrons of $\beta^2/2 < 4.7 \times 10^{-29}$ \cite{bartalucciNewExperimentalLimit2006}, the most stringent constraint at the time. Its successor, VIP-2, pushed this further, establishing limits of $\beta^2/2 \le 6.8 \times 10^{-42} \quad (\text{Bayesian, 90\% CL})$ and $\beta^2/2 \le 7.1 \times 10^{-42} \quad (\text{Frequentist } \mathrm{CL}_s,\,90\%\,\text{CL})$ \cite{napolitanoTestingPauliExclusion2022}. The ongoing effort is extending the search across the periodic table, with high-$Z$ targets such as zirconium, silver, palladium, tin \cite{mantiTestingPauliExclusion2024} and germanium as well.\newline
Reaching such limits requires detector technologies that can operate with high efficiency and stability in the keV range. As an example, Broad Energy Germanium (BEGe) detectors operating at LNGS, combine the features of coaxial and low-energy high-purity germanium designs, covering a wide dynamic range from a few keV up to several MeV \cite{agostiniCharacterizationBroadEnergy2011}. They are particularly suited for rare-event searches because of their excellent resolution and efficiency at low energies \cite{agostiniPulseShapeDiscrimination2013}. Achieving energy thresholds of a few keV is technically demanding, as microphonic noise and electronic instabilities tend to obscure faint signals precisely in this region. Access to this low-energy window is essential: it is where the predicted spectra of different collapse models exhibit their strongest discriminating power \cite{piscicchiaXRayEmissionAtomic2024b}, and it is also the key region of interest for PEP-violation searches, since the normal K$\alpha$ line of germanium lies at 9.2 keV.
In recent years, machine-learning (ML) techniques have emerged as powerful tools for improving pulse-shape analysis \cite{hollDeepLearningBased2019,baccoloMachineLearningassistedTechniques2025} and low-energy event selection in germanium detectors, enabling discrimination between single-site and multi-site interactions \cite{misiaszekImprovingSensitivityBEGebased2018} and identification of event classes relevant for rare-event searches such as neutrinoless double-beta decay \cite{caldwellSignalRecognitionEfficiencies2015}. A practical challenge that remains is the consistent recovery of low-energy signals for BEGe detectors, which carry much of the sensitivity for both PEP and collapse-related measurements \cite{piscicchiaOptimizationBEGeDetector2024a}.\newline
In this work, we report recent advances in the operation of a BEGe detector at LNGS. We developed and validated a ML–based event selection strategy, employing a denoising autoencoder (DAE) and a convolutional neural network to enhance discrimination between normal and anomalous events. These methods were applied to data acquired in 2021, demonstrating improved effective energy threshold and a higher signal-to-background ratio. The paper is organized as follows. Section 2 introduces the ML framework for event selection, with emphasis on the denoising and classification stages. Section 3 presents the performance validation on the 2021 dataset and discusses the improvements in resolution, threshold, and background rejection. Section 4 concludes with a summary and an outlook for future developments within the VIP program and related rare-event searches.
%

%
\section{Event Selection Strategy}
\label{sec:bege_selection}
\begin{figure}[h!]
    \centering
    \includegraphics{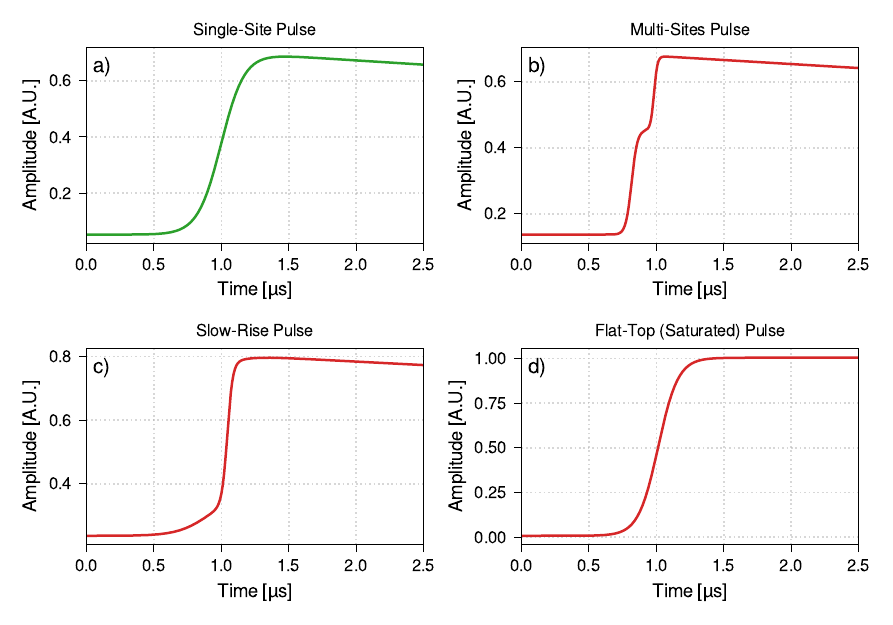}
    \caption{Examples of the four pulse types in the synthetic dataset used to train the DAE: a normal single-site pulse (green) in panel a, and anomalous pulses (red): multi-site (panel b), slow-rise (panel c), and saturated (panel d).}
    \label{fig:pulse_types}
\end{figure}

\begin{figure}[h!]
    \centering
    \includegraphics[width=\linewidth]{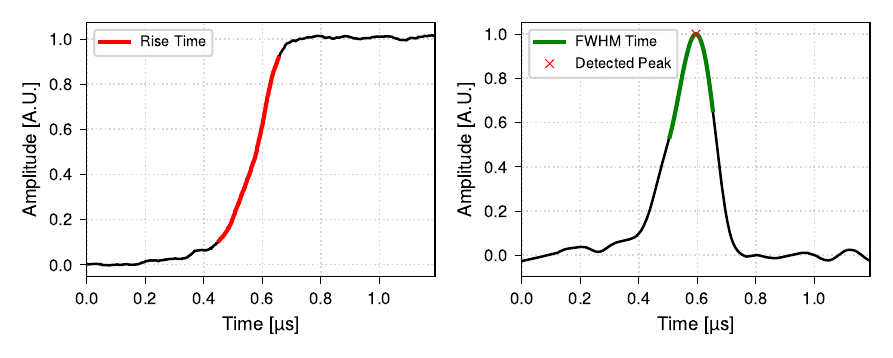}
    \caption{The rise time (left) is defined as the interval between 10\% and 90\% of the waveform amplitude (red). The FWHM time (right) is defined from the derivative of the waveform, with the FWHM interval highlighted in green.}
    \label{fig:rise_fwhm}
\end{figure}

Event selection was performed using pulse shape analysis (PSA) on the acquired waveforms of the BEGe setup. To automate this selection and integrate it in the data acquisition, we employed ML with a DAE \cite{vincentStackedDenoisingAutoencoders2010} with two main objectives. First, to remove noise from the waveforms and improve the selection of low-energy events. Second, to extract features that help in distinguishing normal events from anomalous ones (e.g. multi-site, saturated, and similar events). The latter is important for the creation of a dataset for supervised training of a classifier connected to the output of the DAE, allowing for anomalous detection of events.\newline\noindent
This ML-aided PSA procedure was validated using the data collected in 2021, consisting of more than 20,000 events. For each event, a waveform is recorded by the BEGe detector electronics through a 400 MHz digitizer spanning 1024 samples, with a total acquisition window of 2.55 $\mu$s. Further details on the BEGe apparatus are provided in our previous article \cite{piscicchiaOptimizationBEGeDetector2024a}. We implemented the DAE within the Keras framework (v3.5.0). The DAE, which operates on both the waveform and its derivative, is based on a one-dimensional (1D) convolutional neural network (CNN) with an encoding and decoding structure. The encoder comprises three convolutional layers with progressively decreasing filter sizes (i.e. 64, 32, 16), each followed by the rectified linear unit (ReLU) activation function, and max-pooling operations. The temporal dimension of the waveform in the input layer was reduced to 128 to decrease network complexity and shorten training time. The chosen input size and number of hidden layers represent a balance between model complexity, training efficiency, and the performance of both the DAE and the attached classifier. The decoder mirrors the encoder structure, using upsampling layers and additional convolutional layers with ReLU activation to restore the original waveform, followed by a final convolutional layer with linear activation to produce the reconstructed signal. The DAE is trained using the Adam optimizer \cite{kingma2015adam} with a learning rate of 10$^{-4}$. No hyperparameter tuning was required, since both the DAE and the CNN classifier achieved excellent performance with their initial configurations.\newline\noindent 
%
To train the DAE, we created a synthetic dataset of pulses that reproduce the typical scenarios of events in our BEGe detector setup (see Figure~\ref{fig:pulse_types}).
%
%
A normal event (see Figure~\ref{fig:pulse_types}a) corresponds to single-site energy depositions occurring within the bulk of the germanium detector, where the electric field is uniform and charge carriers are efficiently collected at the electrodes. These events are characterized by a prompt, steep rising edge and an exponential decay, and represent the expected signal. 
Multi-site pulses (see Figure~\ref{fig:pulse_types}b)  are generated to mimic the response to events in which energy is deposited at multiple, spatially separated locations within the detector volume. Such events can arise from, for example, Compton scattering or coincident background interactions. The resulting pulse shapes typically exhibit multiple rising steps or a more complex rise profile, reflecting the superposition of charge collection from each interaction site. 
Slow-rise pulses (see Figure~\ref{fig:pulse_types}c) are included to model interactions occurring near the detector surfaces, where the electric field strength is reduced. In these regions, charge carriers drift more slowly to the electrodes, resulting in a notably slower rise-time compared to bulk events. 
The inclusion of slow-rise pulses in the training set allows the DAE to learn features associated with surface backgrounds, which can be subsequently suppressed through pulse shape discrimination, thereby improving the overall signal-to-noise ratio. 
Flat-top (see Figure~\ref{fig:pulse_types}d) are also incorporated in the dataset. These pulses are characterized by a rapid rise followed by a plateau, which can occur when the detector response saturates, due to high-energy depositions or limitations in the readout electronics. Including flat-top pulses in the training set is crucial to ensure that the DAE can recognize and appropriately process saturated signals, which may otherwise be misidentified or improperly reconstructed. These categories follow the standard classification established in BEGe PSA studies \cite{agostiniCharacterizationBroadEnergy2011,agostiniPulseShapeDiscrimination2013}.\newline\noindent
Each simulated pulse was superimposed with noise to reproduce the conditions of the experimental setup. Two noise components were considered: (i) white Gaussian noise, representing intrinsic electronic fluctuations, and (ii) colored noise generated by integrating a white noise process to emulate the low-frequency microphonic and environmental instabilities observed in the detector. The relative amplitude of the two components was tuned to match the baseline noise spectrum previously characterized for the BEGe setup \cite{piscicchiaOptimizationBEGeDetector2024a}. Two distinct noise levels were applied, yielding a total of 5000 simulated events evenly divided between high- and low-noise conditions. This approach ensured that the DAE was trained across realistic noise environments, thereby improving its robustness when applied to experimental data.\newline\noindent
All events are preprocessed prior to training through baseline subtraction and normalization by their
amplitude. For each pulse, the baseline is first estimated as the mean of the initial 100 samples and
subtracted from the entire waveform to remove DC offsets. The resulting signal is then processed
with a trapezoidal filter \cite{jordanovDigitalSynthesisPulse1994}, and the amplitude is extracted as the maximum absolute value of the
filtered output beyond the filter response region. The baseline-subtracted pulse is then normalized
by this amplitude, resulting in a unit-amplitude, baseline-corrected signal. This standardized
preprocessing pipeline is applied uniformly to both synthetic and experimental datasets to ensure
robust and consistent inputs throughout model training and evaluation. It is worth noting that
without this baseline subtraction and normalization, the DAE fails to accurately reconstruct pulse
shapes.\newline\noindent
The reconstructed pulses produced by the DAE serve as inputs for the binary CNN classifier for the detection of anomalous events. Each event has two input channels: the normalized pulse taken from the DAE output, which needs no further normalization, and the normalized first derivative of the pulse scaled to the range $[0,1]$. Using both the pulse and its first derivative as separate channels, allows the CNN to learn the relation between the pulse shape and small changes, such as multi-site events.\newline\noindent
To train the classifier, we create a dataset of waveforms with labels, by selecting features from the output of the DAE and placing constraints on their distributions. The following features were used:
\begin{itemize}
\item Rise time: the time needed for the waveform to rise from $10\%$ to $90\%$ of its amplitude.
\item Number of peaks in the first derivative of the waveform.
\item FWHM time: the duration of the full width at half maximum of the derivative for events with only one peak in the derivative.
\item The L$1$-norm of the waveform with respect to the average pulse.
\end{itemize}
A visual inspection of the rise time and FWHM time is shown in Figure~\ref{fig:rise_fwhm}.

\begin{figure}[h!]
    \centering
    \includegraphics[width=\linewidth]{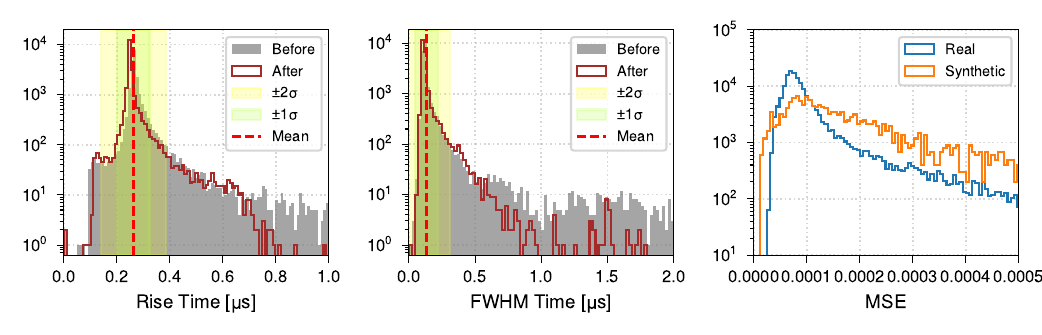}
    \caption{Rise time (left) and FWHM time (center) distributions for the events before and after applying the DAE, with mean and variance intervals. On the right is the mean squared error (MSE) of the DAE on the real and synthetic waveform data.}
    \label{fig:Val_DAE}
\end{figure}

\begin{figure}[h!]
    \centering
    \includegraphics[width=\linewidth]{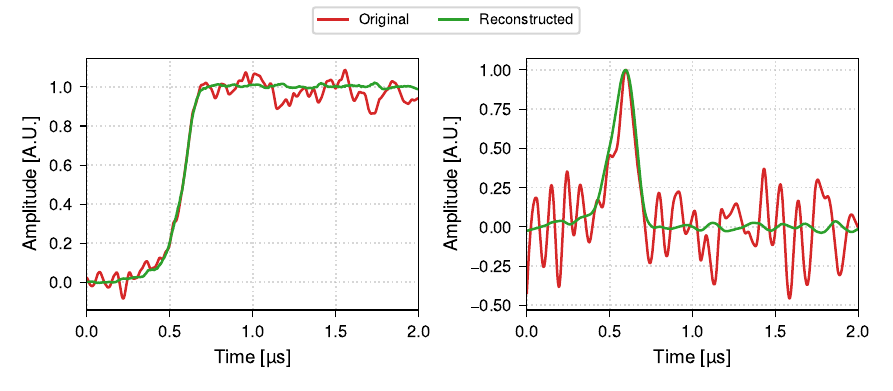}
    \caption{Effect of the DAE on a real waveform (left) and on its derivative (right) for a low-energy pulse which corresponds to an energy of  \textasciitilde15 keV.}
    \label{fig:DAE_recon_example}
\end{figure}

To assess the validity of performing feature extraction on the DAE-processed waveforms, we evaluate the rise time and FWHM for all events before and after denoising, as illustrated in Figure~\ref{fig:Val_DAE}.
%
The DAE does not change the features within $\pm\sigma$ of the main peak. This shows that the cleaning procedure does not distort the waveforms and supports the validation of the DAE transfer from synthetic data to real data. To further support this, we also show in Figure~\ref{fig:Val_DAE} the mean squared error (MSE) of the DAE on real and synthetic data. The DAE gives similar values on both datasets, which indicates that it generalizes well to real data. The larger tail in the synthetic data is due to the different noise level used in the construction of the synthetic dataset. As example, the effect of the DAE on a low energy event, where the reduction of the background is clear, is shown in Figure~\ref{fig:DAE_recon_example}.\newline\noindent
%
Normal events for training are defined as those that satisfy all of the following conditions:
\begin{enumerate}
\item Feature constraints: the rise time, full width at half maximum (FWHM) time, and L1 norm each fall within one standard deviation ($1\sigma$) of their respective distributions (see Figure~\ref{fig:Val_DAE}).
\item Peak count: the event contains exactly one peak in the normalized first derivative, as determined with the \texttt{find\_peaks} function from SciPy \cite{2020SciPy-NMeth}.
\item Labeling: all waveforms are manually inspected and labeled with a custom GUI program and assigned a normal or anomaly label.
\end{enumerate}
The resulting dataset is already well defined by the first two criteria, and manual labeling changes the selection by only $3\%$.
To train the CNN classifier, the dataset was split in training, validation, and testing with proportions 70\%, 15\% and 15\%, respectively. 
The CNN model is composed of sequential 1D convolutional layers with increasing filter sizes (128,64,32) and decreasing kernel widths, each followed by batch normalization and max pooling to extract and condense features from the input. A fully connected dense layer interprets the extracted features, with a dropout layer included to mitigate overfitting. The final output layer uses a sigmoid activation function to produce a probability score for binary classification. The model is trained using the Adam optimizer with a learning rate of 10$^{-5}$ and binary cross-entropy loss, and evaluated using accuracy, precision, and recall metrics. Early stopping of 10 epochs is employed to prevent overfitting and ensure optimal model weights.
\section{Results and Discussion}
\label{sec:results_discussion}
\begin{figure}[h!]
    \centering
    \includegraphics[width=\textwidth]{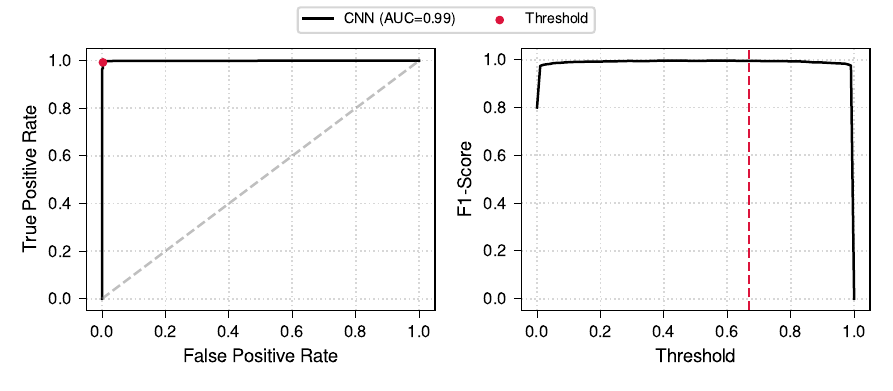}
    \caption{Performance of the CNN classifier applied to DAE-processed waveforms. Left: ROC curve with an AUC of 0.99, with the chosen operating threshold marked in red. Right: accuracy and F1-score as a function of the classification threshold, with the selected operating point indicated by the dashed red line.}
    \label{fig:f1score}
\end{figure}

\begin{figure}[htbp]
    \centering
    \includegraphics[width=\textwidth]{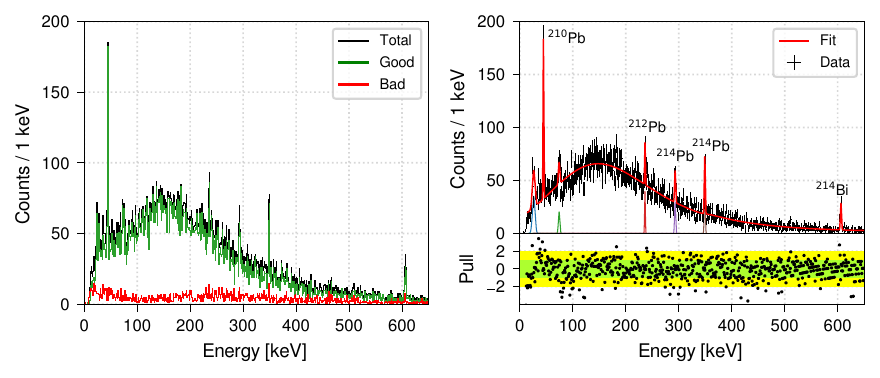}
    \caption{Left: measured spectra after ML-based event selection, showing normal (green), anomalous (red), and total (black) events. Right: fitted spectrum (red) with individual Gaussian components (coloured) and pull distribution below.}
    \label{fig:calib_and_raw_spectrum}
\end{figure}

The DAE--CNN workflow was applied to the 2021 BEGe dataset to evaluate its performance and improvements on the final reconstructed spectrum and assess its suitability for integration into the online data-acquisition (DAQ) of the BEGe for VIP.
After training, the CNN classifier assigns to each event the probability of being a normal (single-site) waveform. The classifier performance is quantified using the receiver operating characteristic (ROC) curve and the F1-score, as shown in Figure~\ref{fig:f1score}, resulting in an area under the curve (AUC) of 0.99. The optimal operating point is defined by the threshold value (0.66) that maximizes the F1-score, corresponding to an accuracy of 95\,\%. At this working point, approximately 13\,\% of events are rejected as anomalous. A feature-based classifier trained on rise time and FWHM also provides good separation power, but computing these features online would introduce significant computational overhead. In contrast, the CNN operates directly on the acquired waveform, maintaining high accuracy while remaining compatible with real-time inference, and allowing the DAQ system to store only classified results rather than full waveforms, thereby substantially reducing the data volume.
%
Events with a classification probability above the optimal threshold are selected as normal (single-site) events. Spectra constructed by selecting events with the classifier can be divided as in Figure~\ref{fig:calib_and_raw_spectrum}. To extract the calibration, a fit with a binned maximum-likelihood approach was done. The calibration uses the four prominent gamma lines from the $^{238}$U decay chain: $^{210}$Pb at 46\,keV, $^{214}$Pb at 295 and 352\,keV, and $^{214}$Bi at 609\,keV. The $^{212}$Pb line at 238 keV is not included in the fit but used as an independent residual test. Each spectrum is fitted with the following model:
\begin{align} f(x) =\;& e^{a_1 + b_1 x}\, \frac{1}{2} \left[1 + \operatorname{erf}\left(\tfrac{x - \mu_1}{\sigma_1}\right)\right] \nonumber + e^{a_2 + b_2 x}\, \frac{1}{2} \left[1 + \operatorname{erf}\left(\tfrac{x - \mu_2}{\sigma_2}\right)\right] \nonumber + \sum_{i} A_i\, \exp\left(-\tfrac{(x - \mu_{g,i})^2}{2\sigma_{g,i}^2}\right). \end{align}
The background is modeled using two error-function components, chosen to be an empirical parameterization that provides a stable and reliable fit for the background. A physically motivated background model will be developed in future studies. The gamma lines are modeled with Gaussian functions. The fitted spectra and pull distributions are shown in Figure~\ref{fig:calib_and_raw_spectrum}. Using the DAE--CNN event selection, the effective energy threshold is lowered to approximately 10\,keV, and the energy resolution at the 46\,keV line is measured to be in the range of 1.0--1.5\,keV (FWHM). The DAE significantly suppresses noise-induced distortions in the sub-15\,keV region, enabling reliable amplitude extraction at low energy.
%
Without the DAE preprocessing, the region below 15\,keV is dominated by electronic noise and microphonic disturbances, limiting the depth of the analysis and reducing the sensitivity to low-energy processes. The DAE reconstructs physically plausible pulse shapes while suppressing noise components, significantly improving amplitude estimation and stabilizing the downstream classification. The ML-based selection enhances the signal-to-background ratio by approximately 14\,\%, with the largest improvement in the 10--50\,keV region where anomalous waveforms contribute substantially to the background. 
Reaching an effective threshold around 10 keV provides full sensitivity for collapse model tests, and with larger future data acquisitions it may become feasible to probe the vicinity of the 9.2 keV K$\alpha$ line relevant for PEP-violation searches in germanium.
Moreover, the DAE preprocessing enables alternative analysis strategies beyond the CNN classifier, including traditional pulse-shape discrimination and feature-based ML methods based on rise time, FWHM, or waveform derivatives. A Python package, \texttt{vipbege} \cite{vipbege}, is provided together with this work, offering tools to extract the waveform features discussed here and to generate simulated waveforms with or without noise. This makes the approach transferable to future BEGe datasets and adaptable to similar low-background detector configurations.
\section{Conclusions}
In this work, we developed and validated a ML–based event-selection strategy for a BEGe detector operated by the VIP collaboration at LNGS. A DAE, trained on synthetic waveforms representative of the main pulse classes, was used to suppress microphonic and electronic noises and to extract robust pulse-shape features. A CNN classifier attached to the DAE output enables fast event selection and anomaly detection.
Applied to the 2021 dataset, the method achieves an AUC of 0.99 and 95\% accuracy, lowers the effective energy threshold from 20 keV to about 10 keV, and improves the signal-to-background ratio by ~14\%. These results demonstrate that ML-based waveform cleaning can significantly enhance the low-energy sensitivity of BEGe detectors, providing a valuable tool for studies of collapse-model signatures and future tests of the Pauli Exclusion Principle in low-background environments.
\section*{Data availability statement}
The labeled dataset of 2021 BEGe waveforms used in this study is publicly available on Zenodo under the DOI: 10.5281/zenodo.17641589. All scripts required to reproduce the figures presented in this article are available at:
\texttt{git@github.com:simonemanti/figures\_bege.git}.
\acknowledgments
%
We thank the Gran Sasso Underground Laboratory of INFN, INFN-LNGS, and its Director, Ezio Previtali, the LNGS staff, and the low radioactivity laboratory for the experimental activities dedicated to high-sensitivity tests of the Pauli exclusion principle.
\section*{Funding}
This publication was made possible through the support of the INFN institute and Centro
Ricerche Enrico Fermi Museo Storico della Fisica e Centro Studi e Ricerche “Enrico Fermi” Institute.
We acknowledge the support of grant 62099 from the John Templeton Foundation. The opinions
expressed in this publication are those of the authors and do not necessarily reflect the views of the
John Templeton Foundation. We acknowledge support from the Foundational Questions Institute and
the Fetzer Franklin Fund, a donor-advised fund of the Silicon Valley Community Foundation (grant
Nos. FQXi-RFP-CPW-2008 and FQXi-MGB-2011), and from the H2020 FET TEQ (grant No. 766900).
We thank the Austrian Science Foundation (FWF), which supports the VIP-2 project with grants
P25529-N20, project P 30635-N36 and W1252-N27 (doctoral college particles and interactions).

\bibliographystyle{JHEP}
\bibliography{BEGe.bib}

\providecommand{\href}[2]{#2}\begingroup\raggedright\begin{thebibliography}{10}

\bibitem{porcelliStrongestConstraintParastatistical2025}
A.~Porcelli, K.~Piscicchia, M.~Bazzi, N.~Bortolotti, M.~Bragadireanu, M.~Cargnelli et~al., \emph{Strongest constraint on the parastatistical {{Quon}} model with the {{VIP-2}} measurements}, \href{https://doi.org/10.1038/s41598-025-25444-z}{\emph{Scientific Reports} {\bfseries 15} (2025) 41544}.

\bibitem{bassiModelsWavefunctionCollapse2013}
A.~Bassi, K.~Lochan, S.~Satin, T.P.~Singh and H.~Ulbricht, \emph{Models of wave-function collapse, underlying theories, and experimental tests}, \href{https://doi.org/10.1103/RevModPhys.85.471}{\emph{Reviews of Modern Physics} {\bfseries 85} (2013) 471}.

\bibitem{pauliConnectionSpinStatistics1940}
W.~Pauli, \emph{The {{Connection Between Spin}} and {{Statistics}}}, \href{https://doi.org/10.1103/PhysRev.58.716}{\emph{Physical Review} {\bfseries 58} (1940) 716}.

\bibitem{dysonStabilityMatter1967}
F.J.~Dyson and A.~Lenard, \emph{Stability of {{Matter}}. {{I}}}, \href{https://doi.org/10.1063/1.1705209}{\emph{Journal of Mathematical Physics} {\bfseries 8} (1967) 423}.

\bibitem{glendenningSpecialGeneralRelativity2010}
N.K.~Glendenning, \emph{Special and {{General Relativity}}: {{With Applications}} to {{White Dwarfs}}, {{Neutron Stars}} and {{Black Holes}}}, Springer Science \& Business Media (Apr., 2010).

\bibitem{streaterPCTSpinStatistics2000}
R.F.~Streater and A.S.~Wightman, \emph{{{PCT}}, {{Spin}} and {{Statistics}}, and {{All}} That}, Princeton University Press (2000).

\bibitem{guidoAlgebraicSpinStatistics1995}
D.~Guido and R.~Longo, \emph{An algebraic spin and statistics theorem}, \href{https://doi.org/10.1007/BF02101806}{\emph{Communications in Mathematical Physics} {\bfseries 172} (1995) 517}.

\bibitem{greenbergParticlesSmallViolations1991}
O.W.~Greenberg, \emph{Particles with small violations of {{Fermi}} or {{Bose}} statistics}, \href{https://doi.org/10.1103/PhysRevD.43.4111}{\emph{Physical Review D} {\bfseries 43} (1991) 4111}.

\bibitem{greenbergTheoriesViolationStatistics2000}
O.W.~Greenberg, \emph{Theories of violation of statistics}, \href{https://doi.org/10.1063/1.1337721}{\emph{AIP Conference Proceedings} {\bfseries 545} (2000) 113}.

\bibitem{balachandranSpinStatisticsGroenewold2006}
A.P.~Balachandran, G.~Mangano, A.~Pinzul and S.~Vaidya, \emph{Spin and statistics on the groenewold--moyal plane: Pauli-forbidden levels and transitions}, \href{https://doi.org/10.1142/S0217751X06031764}{\emph{International Journal of Modern Physics A} {\bfseries 21} (2006) 3111}.

\bibitem{rambergExperimentalLimitSmall1990}
E.~Ramberg and G.A.~Snow, \emph{Experimental limit on a small violation of the {{Pauli}} principle}, \href{https://doi.org/10.1016/0370-2693(90)91762-Z}{\emph{Physics Letters B} {\bfseries 238} (1990) 438}.

\bibitem{ghirardiUnifiedDynamicsMicroscopic1986}
G.C.~Ghirardi, \emph{Unified dynamics for microscopic and macroscopic systems}, \href{https://doi.org/10.1103/PhysRevD.34.470}{\emph{Physical Review D} {\bfseries 34} (1986) 470}.

\bibitem{ghirardiContinuousspontaneousreductionModelInvolving1990}
G.~Ghirardi, R.~Grassi and A.~Rimini, \emph{Continuous-spontaneous-reduction model involving gravity}, \href{https://doi.org/10.1103/PhysRevA.42.1057}{\emph{Physical Review A} {\bfseries 42} (1990) 1057}.

\bibitem{diosiModelsUniversalReduction1989}
L.~Di{\'o}si, \emph{Models for universal reduction of macroscopic quantum fluctuations}, \href{https://doi.org/10.1103/PhysRevA.40.1165}{\emph{Physical Review A} {\bfseries 40} (1989) 1165}.

\bibitem{penroseGravitysRoleQuantum1996}
R.~Penrose, \emph{On {{Gravity}}'s role in {{Quantum State Reduction}}}, \href{https://doi.org/10.1007/BF02105068}{\emph{General Relativity and Gravitation} {\bfseries 28} (1996) 581}.

\bibitem{piscicchiaXRayEmissionAtomic2024b}
K.~Piscicchia, S.~Donadi, S.~Manti, A.~Bassi, M.~Derakhshani, L.~Di{\'o}si et~al., \emph{X-{{Ray Emission}} from {{Atomic Systems Can Distinguish}} between {{Prevailing Dynamical Wave-Function Collapse Models}}}, \href{https://doi.org/10.1103/PhysRevLett.132.250203}{\emph{Physical Review Letters} {\bfseries 132} (2024) 250203}.

\bibitem{belliniCosmicmuonFluxAnnual2012}
G.~Bellini, J.~Benziger, D.~Bick, G.~Bonfini, D.~Bravo, M.B.~Avanzini et~al., \emph{Cosmic-muon flux and annual modulation in {{Borexino}} at 3800 m water-equivalent depth}, \href{https://doi.org/10.1088/1475-7516/2012/05/015}{\emph{Journal of Cosmology and Astroparticle Physics} {\bfseries 2012} (2012) 015}.

\bibitem{xenoncollaborationFirstDarkMatter2023}
{XENON Collaboration}, E.~Aprile, K.~Abe, F.~Agostini, S.~Ahmed~Maouloud, L.~Althueser et~al., \emph{First {{Dark Matter Search}} with {{Nuclear Recoils}} from the {{XENONnT Experiment}}}, \href{https://doi.org/10.1103/PhysRevLett.131.041003}{\emph{Physical Review Letters} {\bfseries 131} (2023) 041003}.

\bibitem{majoranacollaborationSearchNeutrinolessDouble$ensuremathbeta$2018}
{Majorana Collaboration}, C.E.~Aalseth, N.~Abgrall, E.~Aguayo, S.I.~Alvis, M.~Amman et~al., \emph{Search for {{Neutrinoless Double-}}\$\textbackslash ensuremath\textbraceleft\textbackslash beta\textbraceright\$ {{Decay}} in \$\textasciicircum\textbraceleft 76\textbraceright\textbackslash mathrm\textbraceleft{{Ge}}\textbraceright\$ with the {{Majorana Demonstrator}}}, \href{https://doi.org/10.1103/PhysRevLett.120.132502}{\emph{Physical Review Letters} {\bfseries 120} (2018) 132502}.

\bibitem{gerdacollaborationFinalResultsGERDA2020}
{GERDA Collaboration}, M.~Agostini, G.R.~Araujo, A.M.~Bakalyarov, M.~Balata, I.~Barabanov et~al., \emph{Final {{Results}} of {{GERDA}} on the {{Search}} for {{Neutrinoless Double-}}\$\textbackslash ensuremath\textbraceleft\textbackslash beta\textbraceright\$ {{Decay}}}, \href{https://doi.org/10.1103/PhysRevLett.125.252502}{\emph{Physical Review Letters} {\bfseries 125} (2020) 252502}.

\bibitem{ackermannGerdaExperimentSearch2013}
K.-H.~Ackermann, M.~Agostini, M.~Allardt, M.~Altmann, E.~Andreotti, A.M.~Bakalyarov et~al., \emph{The {{Gerda}} experiment for the search of 0{$\nu\beta\beta$} decay in {{76Ge}}}, \href{https://doi.org/10.1140/epjc/s10052-013-2330-0}{\emph{The European Physical Journal C} {\bfseries 73} (2013) 2330}.

\bibitem{bartalucciNewExperimentalLimit2006}
S.~Bartalucci, S.~Bertolucci, M.~Bragadireanu, M.~Cargnelli, M.~Catitti, C.~Curceanu~(Petrascu) et~al., \emph{New experimental limit on the {{Pauli}} exclusion principle violation by electrons}, \href{https://doi.org/10.1016/j.physletb.2006.07.054}{\emph{Physics Letters B} {\bfseries 641} (2006) 18}.

\bibitem{napolitanoTestingPauliExclusion2022}
F.~Napolitano, S.~Bartalucci, S.~Bertolucci, M.~Bazzi, M.~Bragadireanu, C.~Capoccia et~al., \emph{Testing the {{Pauli Exclusion Principle}} with the {{VIP-2 Experiment}}}, \href{https://doi.org/10.3390/sym14050893}{\emph{Symmetry} {\bfseries 14} (2022) 893}.

\bibitem{mantiTestingPauliExclusion2024}
S.~Manti, M.~Bazzi, N.~Bortolotti, C.~Capoccia, M.~Cargnelli, A.~Clozza et~al., \emph{Testing the {{Pauli Exclusion Principle}} across the {{Periodic Table}} with the {{VIP-3 Experiment}}}, \href{https://doi.org/10.3390/e26090752}{\emph{Entropy} {\bfseries 26} (2024) }.

\bibitem{agostiniCharacterizationBroadEnergy2011}
M.~Agostini, E.~Bellotti, R.~Brugnera, C.M.~Cattadori, A.~D'Andragora, A.~di~Vacri et~al., \emph{Characterization of a broad energy germanium detector and application to neutrinoless double beta decay search in {{76Ge}}}, \href{https://doi.org/10.1088/1748-0221/6/04/P04005}{\emph{Journal of Instrumentation} {\bfseries 6} (2011) P04005}.

\bibitem{agostiniPulseShapeDiscrimination2013}
M.~Agostini, M.~Allardt, E.~Andreotti, A.M.~Bakalyarov, M.~Balata, I.~Barabanov et~al., \emph{Pulse shape discrimination for {{Gerda Phase I}} data}, \href{https://doi.org/10.1140/epjc/s10052-013-2583-7}{\emph{The European Physical Journal C} {\bfseries 73} (2013) 2583}.

\bibitem{hollDeepLearningBased2019}
P.~Holl, L.~Hauertmann, B.~Majorovits, O.~Schulz, M.~Schuster and A.J.~Zsigmond, \emph{Deep learning based pulse shape discrimination for germanium detectors}, \href{https://doi.org/10.1140/epjc/s10052-019-6869-2}{\emph{The European Physical Journal C} {\bfseries 79} (2019) 450}.

\bibitem{baccoloMachineLearningassistedTechniques2025}
G.~Baccolo, A.~Barresi, D.~Chiesa, A.~Giachero, D.~Labranca, R.~Moretti et~al., \emph{Machine learning-assisted techniques for {{Compton-background}} discrimination in {{Broad Energy Germanium}} ({{BEGe}}) detector}, \href{https://doi.org/10.1140/epjc/s10052-025-14042-y}{\emph{The European Physical Journal C} {\bfseries 85} (2025) 332}.

\bibitem{misiaszekImprovingSensitivityBEGebased2018}
M.~Misiaszek, K.~Panas, M.~Wojcik, G.~Zuzel and M.~Hult, \emph{Improving sensitivity of a {{BEGe-based}} high-purity germanium spectrometer through pulse shape analysis}, \href{https://doi.org/10.1140/epjc/s10052-018-5852-7}{\emph{The European Physical Journal C} {\bfseries 78} (2018) 392}.

\bibitem{caldwellSignalRecognitionEfficiencies2015}
A.~Caldwell, F.~Cossavella, B.~Majorovits, D.~Palioselitis and O.~Volynets, \emph{Signal recognition efficiencies of artificial neural-network pulse-shape discrimination in {{HPGe}} \$\$\textbackslash varvec\textbraceleft 0\textbackslash nu \textbackslash beta \textbackslash beta \textbraceright\$\$-decay searches}, \href{https://doi.org/10.1140/epjc/s10052-015-3573-8}{\emph{The European Physical Journal C} {\bfseries 75} (2015) 350}.

\bibitem{piscicchiaOptimizationBEGeDetector2024a}
K.~Piscicchia, A.~Clozza, D.L.~Sirghi, M.~Bazzi, N.~Bortolotti, M.~Bragadireanu et~al., \emph{Optimization of a {{BEGe Detector Setup}} for {{Testing Quantum Foundations}} in the {{Underground LNGS Laboratory}}}, \href{https://doi.org/10.3390/condmat9020022}{\emph{Condensed Matter} {\bfseries 9} (2024) 22}.

\bibitem{vincentStackedDenoisingAutoencoders2010}
P.~Vincent, H.~Larochelle, I.~Lajoie, Y.~Bengio and P.-A.~Manzagol, \emph{Stacked {{Denoising Autoencoders}}: {{Learning Useful Representations}} in a {{Deep Network}} with a {{Local Denoising Criterion}}}, {\emph{Journal of Machine Learning Research} {\bfseries 11} (2010) 3371}.

\bibitem{kingma2015adam}
D.P.~Kingma and J.~Ba, \emph{Adam, a method for stochastic optimization},  in \emph{Proceedings of the 3rd International Conference on Learning Representations}, 2015.

\bibitem{jordanovDigitalSynthesisPulse1994}
V.T.~Jordanov and G.F.~Knoll, \emph{Digital synthesis of pulse shapes in real time for high resolution radiation spectroscopy}, \href{https://doi.org/10.1016/0168-9002(94)91011-1}{\emph{Nuclear Instruments and Methods in Physics Research Section A: Accelerators, Spectrometers, Detectors and Associated Equipment} {\bfseries 345} (1994) 337}.

\bibitem{2020SciPy-NMeth}
P.~Virtanen, R.~Gommers, T.E.~Oliphant, M.~Haberland, T.~Reddy, D.~Cournapeau et~al., \emph{{{SciPy}} 1.0: {{Fundamental}} algorithms for scientific computing in python}, \href{https://doi.org/10.1038/s41592-019-0686-2}{\emph{Nature Methods} {\bfseries 17} (2020) 261}.

\bibitem{vipbege}
S.a.S.Y.~Manti, \emph{Vipbege: {{Python}} tools for {{BEGe}} waveform processing},  2025.

\end{thebibliography}\endgroup


\end{document}